\documentclass[aps,prb,twocolumn,showpacs,groupedaddress]{revtex4}
\usepackage{graphicx}

\begin{document}

\title{Strong spin-lattice coupling in multiferroic HoMnO$_{3}$: Thermal
expansion anomalies and pressure effect}
\author{C. dela Cruz$^{1}$, F. Yen$^{1}$, B. Lorenz$^{1}$, Y. Q. Wang$^{1}$, Y. Y. Sun$^{1}$,
M. M. Gospodinov$^{2}$, and C. W. Chu$^{1,3,4}$}
\affiliation{$^{1}$Department of Physics and TCSUH, University of
Houston, Houston, TX 77204-5002}
\affiliation{$^{2}$Institute of
Solid State Physics, Bulgarian Academy of Sciences, 1784 Sofia,
Bulgaria}
\affiliation{$^{3}$Lawrence Berkeley National Laboratory,
1 Cyclotron Road, Berkeley, CA 94720}
\affiliation{$^{4}$Hong Kong
University of Science and Technology, Hong Kong, China}

\begin{abstract}
Evidence for a strong spin-lattice coupling in multiferroic
HoMnO$_3$ is derived from thermal expansion measurements along $a$-
and $c$-axis. The magnetoelastic effect results in sizable anomalies
of the thermal expansivities at the antiferromagnetic ($T_N$) and
the spin rotation ($T_{SR}$) transition temperatures as well as in a
negative $c$-axis expansivity below room temperature. The coupling
between magnetic orders and dielectric properties below $T_N$ is
explained by the lattice strain induced by the magnetoelastic
effect. At $T_{SR}$ various physical quantities show discontinuities
that are thermodynamically consistent with a first order phase
transition.
\end{abstract}

\pacs{65.40.-b, 75.30.Kz, 75.40.-s, 75.50.Ee, 77.80.-e}
\maketitle

\date{\today }











Among the multiferroic materials, the rare-earth manganites have
attracted increasing attention because of a wealth of physical
phenomena related to the coexistence of ferroelectricity with
antiferromagnetic (AFM) orders. The coupling and mutual interference
of ferroelectric (FE) and magnetic orders is of fundamental interest
and bears the potential for future applications. For example, the
orthorhombic $R$MnO$_3$ ($R$=Eu to Dy) undergo several magnetic
phase transitions accompanied by distinct dielectric
anomalies.\cite{1,2,3,4} Ferroelectricity and large
magneto-dielectric coupling have been observed in some of the
compounds.\cite{1,3}

The $R$MnO$_3$ with smaller
rare-earth ions ($R$ from Ho to Lu, and Y) crystallize in the hexagonal $%
P6_{3}cm$ structure with ferroelectricity arising well above room
temperature. The onset of AFM order of the Mn-spins below 100 K
gives rise to interesting physical effects related to the coupling
of both orders. The first signature of the magneto-dielectric effect
in hexagonal manganites was discovered in YMnO$_3$ as an anomaly of
the dielectric constant at $T_{N}\approx$70 K.\cite{5} Similar
anomalies have been subsequently reported for almost all hexagonal
$R$MnO$_3$.\cite{6} The magnetic order of the Mn$^{3+}$ spins is
geometrically frustrated since the Mn ions form a triangular lattice
in the $a$-$b$ plane. Additional phase transitions at temperatures
below 10 K are observed in some hexagonal $R$MnO$_3$ with magnetic
$R^{3+}$ due to the $R$-$R$ exchange correlations. Changes in the
magnetic structure of the Mn-spins have been reported for $R$=Lu,
Sc, Ho at intermediate temperatures.\cite{10} The interactions
between the FE polarization, the frustrated AFM order of the
Mn-spins, and the $R$-ion magnetic moments give rise to a complex
magnetic phase diagram as was recently revealed for example, in
HoMnO$_3$.\cite{7,8} At zero magnetic field the AFM transition in
HoMnO$_3$ at $T_N$=76 K is followed by a Mn-spin rotation transition
with the onset of AFM order of the Ho moments at $T_{SR}$=33 K and
another magnetic transition at $T_2$=5.2 K characterized by a
substantial increase of the Ho sublattice magnetization combined
with another rotation of Mn spins.\cite{9} Optical\cite{10} and
neutron scattering experiments\cite{9} have identified the magnetic
symmetry of the phase between $T_{SR}$ and $T_N$ as
$P\underline{6}_{3}\underline{c}m$ (with Mn-spins perpendicular to
the
hexagonal $a$-axis) and between $T_2$ and $T_{SR}$ as $P\underline{6}_{3}c%
\underline{m}$ (Mn-spins rotated by 90$^\circ$ with respect to the $P%
\underline{6}_{3}\underline{c}m$ phase). All three magnetic phase
changes are accompanied by distinct anomalies of the dielectric
constant $\varepsilon$, most notably a very sharp peak of
$\varepsilon$ at $T_{SR}$ that was discovered very
recently.\cite{11}

The dielectric anomalies at the magnetic transitions of HoMnO$_3$
evidence a strong correlation of the magnetic and FE orders. Whereas
the direct coupling between the in-plane magnetic moments of the
Mn-ions and the $c$-axis FE polarization is not allowed for magnetic
symmetries $P\underline{6}_{3}\underline{c}m$ and
$P\underline{6}_{3}c\underline{m}$\cite{12} an indirect coupling via
magnetoelastic deformation and lattice strain was proposed to
account for the observations.\cite{11} No clear evidence for lattice
distortions or strain has been reported so far.

We have therefore measured the thermal expansion coefficients
$\alpha$ along $a$- and $c$-axis of HoMnO$_3$ over a large
temperature range. We find distinct anomalies of $\alpha_a$ and
$\alpha_c$ at $T_N $ and at $T_{SR}$ and a negative c-axis
expansivity at all $T$ below room temperature revealing
extraordinarily strong magnetic correlation and spin-lattice
coupling effects. A sudden increase of the volume at $T_{SR}$
suggests the first order nature of the spin rotation transition that
is confirmed by the thermodynamic consistency of the volume,
magnetization, and entropy discontinuities across the transition as
well as the pressure and magnetic field dependence of $T_{SR}$.

Single crystals of HoMnO$_3$ have been grown from the flux\cite%
{8} and by the floating zone method. The linear thermal expansivity
was measured over a large temperature range below 300 K employing
the strain-gage method. Below 100 K a high-precision capacitance
dilatometer was used to resolve the thermal expansion anomalies near
the magnetic phase transition temperatures, $T_N$ and $T_{SR}$. The
pressure dependence of $T_{SR}$ was investigated by monitoring the
sharp peak of the dielectric constant\cite{11} for pressures up to
1.7 GPa in a Be-Cu high pressure clamp.

The results of the dilatometric measurements of HoMnO$%
_3$ are summarized in Fig. 1. Whereas the $a$-axis behaves 'normal'
and shrinks with decreasing $T$, the $c$-axis length steadily
increases from room temperature to lower $T$. This unusual behavior
indicates a strong magnetic exchange and spin-lattice coupling of
the Mn-spins and will be discussed later. At $T_{N}$ the linear
expansivities $\alpha_a$ and $\alpha_c$ exhibit distinct
$\lambda$-type anomalies (left inset of Fig. 1) with opposite signs.
In cooling through $T_N$ the in-plane distances are reduced and the
$c $-axis expands resulting in an abrupt change of slope of $a(T)$
and $c(T)$ in opposite directions. The $\lambda$-shape of the peaks
of $\alpha_a$ and $\alpha_c$ is typical for a second
order phase transition with a broad critical region. A similarly pronounced $%
\lambda$-type anomaly was also observed in the specific heat,
$C_{p}(T)$, of HoMnO$_3$ at $T_N$.\cite{8,14} The strong anomalies
of $\alpha$ and $C_p$ are evidence for a large magnetoelastic
coupling. The Mn-spins are strongly correlated via the AFM
superexchange interaction in the hexagonal $a$-$b$ plane. The
magnetic exchange coupling along the $c$-axis is much weaker. The
large in-plane exchange interaction should stabilize AFM order at
relatively high temperature. However, due to geometric frustration
of the Mn-spins the magnetic
phase transition takes place at much lower temperature (%
$T_N$=76 K) with a special alignment of the magnetic moments so that
neighboring Mn-spins form an angle of 120$^\circ$.\cite{7} The small
entropy change associated with the AFM transition (only 10 to 15 \%
of the maximum value of Rln5)\cite{8,14} is an indication of the
existence of sizable short-range correlations between the Mn-spins
above $T_N$, as derived from magnetic and neutron scattering
data.\cite{23} Strong magnetic correlations as well as long-range
order are known to be a common origin of lattice strain when the
magnetic energy of the system of interacting moments is lowered by a
change of the interatomic distances.\cite{15,16} In the anisotropic
structure of HoMnO$_3$ the major effect of the Mn-spin exchange
correlations is a reduction of the in-plane distances resulting in a
magnetic contribution to the thermal expansion and an enhancement of
$\alpha _{a}$. The shortening of the $a$-axis at $T_N$ increases the
force constants between the ions and provides a natural explanation
for the abnormal increase of the frequency of some in-plane phonons
recently observed below $T_N$ in HoMnO$_3$.\cite{17} The existence
of strong spin-lattice coupling in hexagonal $R$MnO$_3$ was also
concluded very recently from the observation of a large suppression
of the thermal conductivity in a broad temperature range above $T_N$
in YMnO$_3$ and HoMnO$_3$.\cite{21}

\begin{figure}[tbp]
\includegraphics[width=2.5in]{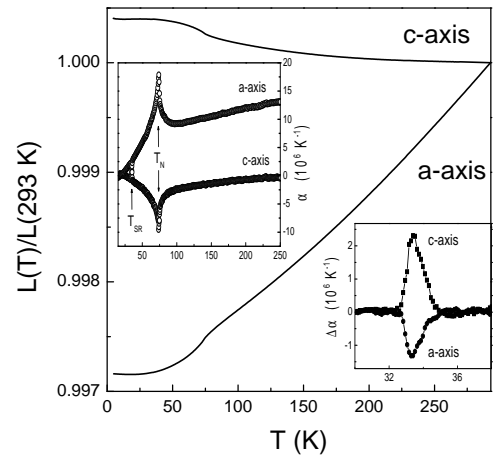}
\caption{Temperature dependence of $a$- and $c$-axis of HoMnO$_3$
(main panel). Left inset: $\alpha_a$ and $\alpha_c$ with distinct
anomalies at $T_N$ and $T_{SR}$. Right inset: Anomalies of
$\alpha_a$ and $\alpha_c$ at $T_{SR}$ (the lattice contribution was
subtracted).}
\end{figure}
The negative expansivity of the $c$-axis below room temperature
(Fig. 1) is observed for the first time in hexagonal $R$MnO$_3$ and
its origin is yet to be explored. For anisotropic compounds it is
not unusual that the thermal expansion in one
crystallographic direction is negative within a limited temperature range.%
\cite{18} In HoMnO$_3$, however, $\alpha _c$ is negative at all $T$
below room temperature. Negative values of $\alpha$ can originate
either from low-energy transverse acoustic (TA) modes of vibration,
as for example in crystals of Si, Ga, CuCl, and others, or from
magnetic interactions with strong spin-lattice coupling.\cite{18}
The soft TA modes are common in open crystal structures of low
coordination number and may lead to unidirectional or isotropic
negative expansivities as, for example, observed in
$\beta$-LiAlSiO$_4$ and ZrW$_2$O$_8$.\cite{19,20} However, the
hexagonal rare earth manganites do not meet the conditions of low
coordination and soft TA phonons have not been observed in the
compounds. We therefore attribute the negative $c$-axis expansivity
to the strong magnetic correlations and magnetoelastic coupling in
the compound. The magnetic correlations of the Mn-spins are
strongest in the $a$-$b$ plane and they increase with decreasing
temperature resulting in a magnetic contribution to the in-plane
thermal expansivities. At the same time, mediated by elastic forces
of the lattice, the $c$-axis expands with decreasing temperature, as
observed in our experiments (Fig. 1). This negative expansion effect
obviously dominates over the commonly positive contribution to the
$c $-axis expansion that is due to the lattice anharmonicities. The
volume expansivity is positive over the whole temperature range.
These effects are particularly strong close to the AFM transition as
reflected in the peaks of $\alpha_a$ and $\alpha_c$ at $T_N$ (inset
of Fig. 1). The opposite directions of both peaks provide further
support to our conclusion. The magnetic moments of the Ho$^{3+}$
ions order at lower temperatures (below $T_{SR}$) and their effect
on the lattice strain between $T_N$ and room temperature can be
considered to be small.

The physical origin of the coupling between the magnetic order and
dielectric properties observed in most of the hexagonal $R$MnO$_3$
is not yet understood.\cite{5,6,11} In
$P\underline{6}_{3}\underline{c}m$ and
$P\underline{6}_{3}c\underline{m}$ symmetries the direct coupling
between the in-plane staggered magnetization and
the $c$-axis ferroelectric polarization is not allowed.\cite%
{11,12} Therefore, the dielectric anomalies at $T_N$ have to be a
second order effect, possibly mediated by lattice strain. Our
thermal expansion measurements provide the first direct evidence for
the existence of a sizable lattice distortion close to $T_N$. This
distortion affects the temperature dependence of $\varepsilon (T)$
of HoMnO$_3$. The magneto-dielectric effect at and below $T_N$ was
recently described by a model that includes the AFM Heisenberg
exchange interaction, a double well potential for the lattice
displacements giving rise to ferroelectricity, and a spin-phonon
interaction term.\cite{22} For certain values of the spin-lattice
coupling constant the drop of $\varepsilon(T)$ below $T_N$ could be
qualitatively reproduced. Thereby, $\varepsilon(T)$ is a function of
the inverse square of the FE displacement (along c-axis). To
qualitatively verify this correlation we compare $\varepsilon(T)$
with $c(T)^{-2}$ in Fig. 2. The perfect scaling of both quantities
over a broad temperature range proves unambiguously that the
dielectric properties and the lattice strain induced by the magnetic
correlations are intimately related.

The peak of $\varepsilon$ at $T_{SR}$ was shown by us to be
associated with the spin rotation transition via an intermediate
phase\cite{11} and its existence was confirmed very
recently.\cite{21} This sharp enhancement of $\varepsilon(T_{SR})$
was attributed to arise from a contribution of magnetic domain walls
of domains with $P\underline{6}_3$ symmetry and the allowed linear
magnetoelectric effect in the walls.\cite{24} The step-like increase
of $\varepsilon$ at 5 K was observed before\cite{6} and it is
related to another major change of the magnetic order involving the
Ho moments.\cite{7,9} The possible magnetoelectric interactions
below $T_{SR}$ and its microscopic origin have been discussed
recently including the effects of asymmetric Dzyaloshinskii-Moriya
exchange interactions between Ho and Mn moments.\cite{25} The
microscopic interactions between Mn spins, Ho moments, and the FE
order result in the complex phase diagram\cite{8} and interesting
physical effects such as electric field induced ferromagnetic order
etc.\cite{25}

\begin{figure}[tbp]
\includegraphics[width=2.5in]{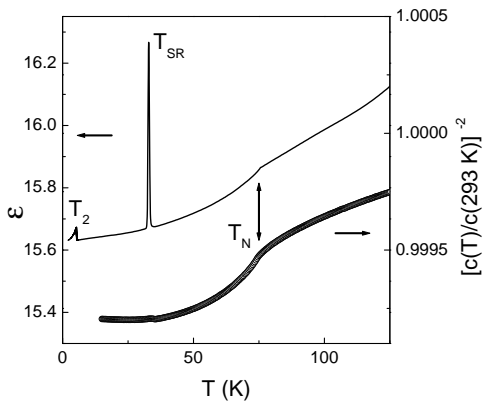}
\caption{Comparison of $\varepsilon (T)$ and the inverse square of
$c(T)$.}
\end{figure}

The spin-rotation transition is very sharp and several dielectric, magnetic,
and thermodynamic quantities change abruptly at $T_{SR}$.\cite{8,11} The
thermal expansion measurements reveal small, but distinctive anomalies at $%
T_{SR}$ resulting in sharp peaks of $\alpha _a$ and $\alpha _c$ with
opposite signs (right inset of Fig. 1). Upon cooling through
$T_{SR}$ the $a$-axis expands by $\Delta a/a$=1.52*10$^{-6}$ and the
$c$-axis shrinks by $\Delta c/c$=-2.44*10$^{-6}$. This behavior is
opposite to the $a$-contraction and $c$-expansion observed at
$T_{N}$. However, it is consistent with the proposed onset of the
AFM order of some Ho-moments oriented along the
$c$-axis\cite{8,9,14} and the expected $c$-axis contraction due to
magnetostrictive effects. It is interesting to
note that the volume below $T_{SR}$ is larger than above with $\Delta V/V$%
=-0.6*10$^{-6}$ (here $\Delta V=V(T>T_{SR})-V(T<T_{SR}$)). The width
of the spin-rotation transition as derived from anomalies of various
quantities is less than 0.6 K.\cite{8} This leads us to suggest the
first order nature of this phase transformation. At first order
transitions, different thermodynamic quantities such as volume and
magnetization exhibit discontinuities that contribute to the total
change of entropy at the transition temperature. The entropy change
at $T_{SR}$ is given by
\begin{equation}
\Delta S=\Delta V\frac{dp}{dT_{SR}}-\frac{1}{2}\frac{\Delta M}{B}\frac{%
d(B^{2})}{dT_{SR}}
\end{equation}
$p$, $B$, $\Delta V$, and $\Delta M$ are pressure, magnetic
induction, volume discontinuity, and magnetization jump at $T_{SR}$,
respectively. The various quantities entering equation (1) are
experimentally accessible and will be used to prove the first order
nature of the spin-rotation transition.
\begin{figure}[tbp]
\includegraphics[angle=-90,width=2.5in]{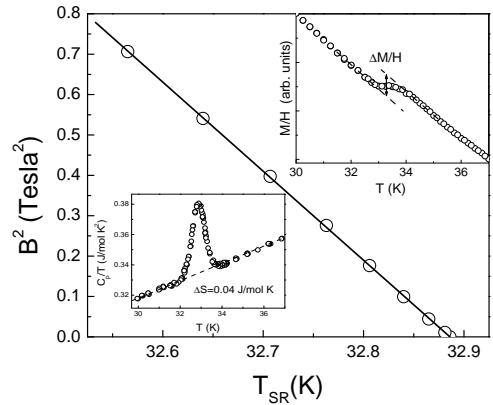}
\caption{Magnetic field dependence of $T_{SR}$ below 1 Tesla. The right and
left insets show the abrupt change of magnetic susceptibility and the heat
capacity peak at $T_{SR}$, respectively.}
\end{figure}
The phase diagram of HoMnO$_{3}$ shows a non-linear decrease of $%
T_{SR}$ with increasing field $B$.\cite{7,8,11} For $B<$1 Tesla we
find a perfect quadratic dependence $T_{SR}\propto B^{2}$ (Fig. 3),
with the slope $d(B^{2})/dT_{SR}$=-2.2 Tesla$^{2}$/K. The jump of
the magnetic susceptibility is determined from dc magnetization
measurements (right inset in Fig. 3) as $\Delta M/B$=818 (Am)/(Vs).
The magnetic contribution to $\Delta S$ in (1) is therefore 0.034
J/(mol K).
\begin{figure}[tbp]
\includegraphics[angle=-90,width=2.5in]{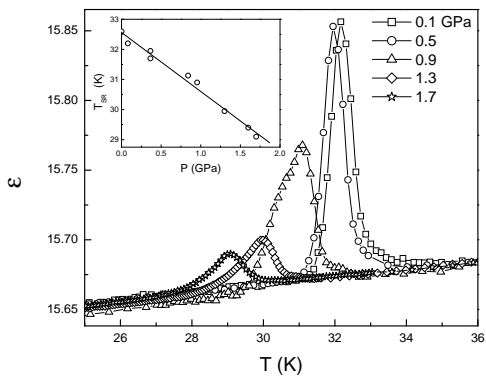}
\caption{Pressure effect on the peak of $\varepsilon$ at $T_{SR}$.
Inset: Pressure dependence of $T_{SR}$.}
\end{figure}
For estimating $\Delta V$d$p$/d$T_{SR}$ the pressure dependence of
$T_{SR}$ needs to be
known. We have measured the pressure shift of the sharp dielectric peak at $%
T_{SR}$ in order to derive $dT_{SR}/dp$ (Fig. 4). The peak
temperature of $\varepsilon (T)$
decreases linearly with applied pressure (inset of Fig. 4) at a rate of $%
dT_{SR}/dp$=-2.05 K/GPa. The decrease of $T_{SR}$ with $p$ is intimately
related to the increase of volume right below $T_{SR}$. The pressure-induced
compression favors the $P\underline{6}_{3}\underline{c}m$ phase (stable
above $T_{SR}$) with the smaller volume on the expense of the $P\underline{6}%
_{3}c\underline{m}$ phase. From the pressure coefficient of $%
T_{SR}$ and the volume change across the transition the mechanical
work contribution to $\Delta S$ is estimated as $\Delta Vdp/dT_{SR}$
= 0.01 J/(mol K). The total entropy change according to (1) is
therefore of the order of 0.044 J/(mol K). This value is to be
compared with $\Delta S$ calculated by integrating the excess
specific heat, $C_{p}/T$, across the phase transition. The small
peak of $C_{p}/T$ at $T_{SR}$ (left inset of Fig. 3) was resolved
only recently\cite{8,21} and it corresponds to an entropy change of
$\Delta S$=0.040 J/(mol K). This value is in very good agreement
with the sum of the two contributions estimated above, i.e. equation
(1) is fulfilled within the experimental uncertainties. This shows
the thermodynamic consistency of all the measured quantities
(specific heat, magnetization, volume expansivity, T-H and T-p phase
boundaries) and it proves the first order nature of the
spin-rotation phase transition in HoMnO$_{3}$. The major
contribution to the entropy change is due to the change of magnetic
order at $T_{SR}$. A detailed analysis of the $c$-axis magnetization
above and below $T_{SR}$ led us to suggest a partial AFM order of
the Ho magnetic moments at $T_{SR}$ and the possible existence of a
correlation between the onset of the Ho order and the Mn-spin
rotation.\cite{8} The small magnetic contribution to $\Delta S$ of
only 0.034 J/(mol K) is consistent with a small sublattice
magnetization deduced from neutron scattering data\cite{9} or a
partial magnetic order at $T_{SR}$ involving only some of the
Ho-moments.\cite{14}

In summary, we demonstrated the existence of extraordinarily strong
spin-spin and spin-lattice couplings over a broad temperature range
in HoMnO$_3$ resulting in a sizable magnetic contribution to the
$a$-axis thermal expansion coefficient and, via elastic coupling, in
the negative $c$-axis expansivity below room temperature. We
conclude that the dielectric anomalies observed in the hexagonal
$R$MnO$_3$ at $T_N$ are a consequence of this spin-lattice coupling.
At the spin-rotation transition of HoMnO$_3$ various physical
quantities show discontinuities indicative of a first order phase
transition. We separate the mechanical (volume expansion) and
magnetic contributions to the total entropy change at $T_{SR}$ and
prove that the entropy balance required by the thermodynamics of
first order transitions is fulfilled.


\begin{acknowledgments}
This work is supported in part by NSF Grant No. DMR-9804325, the
T.L.L. Temple Foundation, the John J. and Rebecca Moores Endowment,
and the State of Texas through the TCSUH and at Lawrence Berkeley
Laboratory by the Director, Office of Energy Research, Office of
Basic Energy Sciences, Division of Materials Sciences of the U.S.
Department of Energy under Contract No. DE-AC03-76SF00098. The work
of M. M. G. is supported by the Bulgarian Science Fund, Grant No.
F-1207.
\end{acknowledgments}


\end{document}